\documentclass[12pt]{article}

\usepackage{amsmath}
\usepackage{graphicx}
\usepackage{cite}

\newcommand{\ket}[1]{| #1 \rangle}
\newcommand{\bra}[1]{\langle #1 |}
\newcommand{\hcs}[1]{#1^\dagger #1}
\newcommand{\expv}[1]{\langle #1 \rangle}
\begin{document}

\begin{center}
{\Large\bf Derivative of the disturbance with respect to information
from quantum measurements}
\vskip .6 cm
Hiroaki Terashima
\vskip .4 cm
{\it Department of Physics, Faculty of Education, Gunma University, \\
Maebashi, Gunma 371-8510, Japan}
\vskip .6 cm
\end{center}

\begin{abstract}
To study the trade-off between information and disturbance,
we obtain the first and second derivatives of
the disturbance with respect to information
for a fundamental class of quantum measurements.
We focus on measurements lying on
the boundaries of the physically allowed regions
in four information--disturbance planes,
using the derivatives
to investigate the slopes and curvatures of these boundaries
and hence clarify the shapes of the allowed regions.
\end{abstract}

\begin{flushleft}
{\footnotesize
}
\end{flushleft}

\section{Introduction}
In quantum theory,
any measurement that provides information about a physical system
also inevitably disturbs the system's state
in a way that depends on the measurement's outcome.
This trade-off between information and disturbance
is of great interest in establishing the foundations of quantum mechanics
and plays an important role in quantum information processing
and communication~\cite{NieChu00} techniques,
such as quantum cryptography~\cite{BenBra84,Ekert91,Bennet92,BeBrMe92}.
Many authors~\cite{FucPer96,Banasz01,FucJac01,BanDev01,Barnum02,%
DArian03,Ozawa04,GenPar05,MiFiFi05,Maccon06,Sacchi06,BusSac06,Banasz06,%
BuHaHo07,CheLee12,RenFan14,FGNZ15,ShKuUe16}
have therefore discussed this trade-off,
using several different formulations.
For example,
Banaszek~\cite{Banasz01} found an inequality
between the amount of information gained
and the size of the state change,
whereas Cheong and Lee~\cite{CheLee12} found one
between the amount of information gained
and the reversibility of the state change.
These inequalities have both been
verified~\cite{SRDFM06,BaChKi08,CZXTLX14,LRHLK14}
in single-photon experiments.

Recently, we have also studied this trade-off,
deriving the allowed regions
in four types of information--disturbance plane~\cite{Terash16}.
These four information--disturbance pairs combine
one information measure, namely
the Shannon entropy~\cite{FucPer96}
or estimation fidelity~\cite{Banasz01},
with one disturbance measure, namely
the operation fidelity~\cite{Banasz01}
or physical reversibility~\cite{KoaUed99}.
The boundaries of the allowed regions
give upper and lower bounds on the information
for a given disturbance,
together with the optimal measurements that saturate the upper bounds.
The optimal measurements are different for each of the four pairs,
because the allowed regions' upper boundaries
have different curvatures
on each of the information--disturbance planes~\cite{Terash16}.

Contrary to expectations,
the allowed regions show that
measurements providing more information
do not necessarily cause larger disturbances.
This is because the allowed regions have finite areas,
i.e., for any given measurement corresponding to an interior point
of an allowed region, there always exists
another measurement that provides more information
with smaller disturbance near that point.
However, measurements that lie on the boundary of an allowed region
in the information--disturbance plane are subject to a trade-off.
Meaning that, modifying them
to increase the information obtained by moving along the boundary
also increases the disturbance according to the boundary's slope.

In this paper,
we obtain the first and second derivatives of
the disturbance with respect to the information
obtained from measurements lying on
the allowed regions' boundaries
for each of the four information--disturbance pairs.
These measurements are described by a diagonal operator
with a continuous parameter, and
applied to a $d$-level system in a completely unknown state.
For such measurements,
we calculate these derivatives to demonstrate
the slopes and curvatures of the allowed regions' boundaries,
clarifying the regions' shapes
and hence, broadening our perspective on the trade-off
between information and disturbance in quantum measurements.
In fact,
it was difficult to judge from
the allowed regions shown in Ref.~\cite{Terash16}
whether the slopes of the boundaries are finite
and whether the curvatures of the boundaries
are negative at some points.
In contrast, the first and second derivatives
obtained in this paper give
the values of the slopes and curvatures of the boundaries
to answer these questions.

The rest of this paper is organized as follows.
Section~\ref{sec:formulation} reviews the procedure
for quantifying the information and the disturbance in quantum measurements,
giving their explicit forms
for a fundamental class of measurements as functions of a certain parameter.
Section~\ref{sec:derivativePara} presents
the first and second derivatives of
the information and the disturbance for such measurements
with respect to this parameter,
while Section~\ref{sec:derivativeInfo} gives
the first and second derivatives of
the disturbance with respect to the information.
Finally, Section~\ref{sec:conclude} summarizes our results.

\section{\label{sec:formulation}Information and Disturbance}
In this section, we recall
the information and the disturbance
in quantum measurements
at the single-outcome
level~\cite{DArian03,Terash10,Terash11,Terash11b,Terash15}
and summarize the results of Ref.~\cite{Terash16}
in order for this paper to be self-contained.
Suppose we want to measure a $d$-level system
that is known to be in one of a predefined set of
pure states $\{\ket{\psi(a)}\}$,
the probability of the system being in the state $\ket{\psi(a)}$
is given by $p(a)$,
but we do not know the actual states of the system.
To study the case where no prior information about the system is available,
we assume that the set $\{\ket{\psi(a)}\}$ consists
of all possible pure states, and $p(a)$ is uniform
according to a normalized invariant measure over the pure states.

First, we quantify the amount of information
provided by a given quantum measurement~\cite{Terash16}.
An ideal quantum measurement~\cite{NieCav97} can be described by a set of
measurement operators $\{\hat{M}_m\}$~\cite{NieChu00} that satisfy
\begin{equation}
\sum_m\hcs{\hat{M}_m}=\hat{I},
\end{equation}
where $m$ denotes the outcome of the measurement
and $\hat{I}$ is the identity operator.
When the system is in state $\ket{\psi(a)}$,
a measurement $\{\hat{M}_m\}$ yields the outcome $m$ with probability
\begin{equation}
p(m|a)=\bra{\psi(a)}\hcs{\hat{M}_m}\ket{\psi(a)}
\end{equation}
and changes the state to
\begin{equation}
\ket{\psi(m,a)}=\frac{1}{\sqrt{p(m|a)}}\,\hat{M}_m\ket{\psi(a)}.
\label{eq:postState}
\end{equation}

The measurement outcome
provides some information about the system's state.
For example, given the outcome $m$,
the probability that the initial state was $\ket{\psi(a)}$
is given by
\begin{equation}
 p(a|m) =\frac{p(m|a)\,p(a)}{p(m)}
\end{equation}
using Bayes's rule, where
\begin{equation}
 p(m) =\sum_a p(m|a)\,p(a)
\end{equation}
is the total probability of the outcome $m$.
This therefore changes
the state probability distribution
from $\{p(a)\}$ to $\{p(a|m)\}$,
decreasing the Shannon entropy by
\begin{align}
  I(m)  &=\left[-\sum_a p(a)\log_2 p(a)\right] \notag \\
        &   \qquad{}-\left[-\sum_a p(a|m)\log_2 p(a|m)\right].
\label{eq:defInformation}
\end{align}
This entropy change, $I(m)$, quantifies
the amount of information provided by
a measurement $\{\hat{M}_m\}$ with outcome $m$~\cite{DArian03,TerUed07b},
and satisfies
\begin{equation}
 0\le I(m) \le\log_2d-\frac{1}{\ln2}[\eta(d)- 1],
\end{equation}
where
\begin{equation}
 \eta(n)= 
\begin{cases}
       \sum^{n}_{k=1}\frac{1}{k} & \mbox{(if $n=1,2,\ldots$)} \\
        0 & \mbox{(if $n=0$)}.
    \end{cases}
\end{equation}
Note that $I(m)$ is a measure of the information
generated by a single outcome,
unlike
\begin{equation}
 I=\sum_m p(m)\,I(m),
\label{eq:avgI}
\end{equation}
which was discussed in Ref.~\cite{FucPer96}.

The measurement outcome $m$ can also be used to
estimate the system's state as $\ket{\varphi(m)}$,
where an optimal $\ket{\varphi(m)}$
is the eigenvector of $\hcs{\hat{M}_m}$
corresponding to its maximum eigenvalue~\cite{Banasz01}.
The quality of this estimate can be evaluated
in terms of the estimation fidelity $G(m)$:
\begin{equation}
 G(m) =\sum_a p(a|m)\,\bigl|\expv{\varphi(m)|\psi(a)}\bigr|^2.
\label{eq:defEstimation}
\end{equation}
This also quantifies the amount of information
provided by the outcome $m$, and satisfies
\begin{equation}
  \frac{1}{d}\le G(m) \le \frac{2}{d+1}.
\end{equation}
Again, note that $G(m)$ relates to a single outcome,
unlike
\begin{equation}
G=\sum_m p(m)\,G(m),
\label{eq:avgG}
\end{equation}
which was discussed in Ref.~\cite{Banasz01}.

Next, we quantify the degree of disturbance
caused by the measurement $\{\hat{M}_m\}$~\cite{Terash16}.
The outcome $m$ changes
the system's state from $\ket{\psi(a)}$ to $\ket{\psi(m,a)}$,
given by Eq.~(\ref{eq:postState}).
The size of this change can be evaluated
using the operation fidelity $F(m)$:
\begin{equation}
 F(m) =\sum_a p(a|m)\bigl|\expv{\psi(a)|\psi(m,a)}\bigr|^2.
\label{eq:defFidelity}
\end{equation}
This quantifies the degree of disturbance caused
when a measurement $\{\hat{M}_m\}$ yields the outcome $m$,
and satisfies
\begin{equation}
 \frac{2}{d+1}\le F(m) \le 1.
\end{equation}
Again, note that $F(m)$ relates to a single outcome,
unlike
\begin{equation}
F=\sum_m p(m)\,F(m),
\label{eq:avgF}
\end{equation}
which was discussed in Ref.~\cite{Banasz01}.

In addition to the size of the state change,
the reversibility of the change can also be used to
quantify the disturbance in the context of physically
reversible measurements~\cite{UedKit92,Imamog93,Royer94,%
UeImNa96,Ueda97,TerUed05,KorJor06,SuAlZu09,XuZho10,KNABHL08,KCRK09}.
Even though $\ket{\psi(a)}$ and $\ket{\psi(m,a)}$ are unknown,
the change can be physically reversed
by a reversing measurement on $\ket{\psi(m,a)}$
if $\hat{M}_m$ has a bounded left
inverse $\hat{M}_m^{-1}$~\cite{UeImNa96,Ueda97}.
Such a reversing measurement can be described by
another set of measurement operators $\{\hat{R}_\mu^{(m)}\}$ that satisfy
\begin{equation}
\sum_\mu\hat{R}^{(m)\dagger}_\mu\hat{R}^{(m)}_\mu=\hat{I}
\end{equation}
and $\hat{R}^{(m)}_{\mu_0}\propto \hat{M}_m^{-1}$
for a particular $\mu=\mu_0$,
where $\mu$ denotes the reversing measurement's outcome.
When this measurement on $\ket{\psi(m,a)}$
yields the preferred outcome $\mu_0$,
the system's state returns to $\ket{\psi(a)}$
because $\hat{R}_{\mu_0}^{(m)}\hat{M}_m\propto\hat{I}$.
The state recovery probability for
an optimal reversing measurement~\cite{KoaUed99} is
\begin{equation}
 R(m,a)=\frac{\inf_{\ket{\psi}}\, \bra{\psi}\hcs{\hat{M}_m}\ket{\psi}}{p(m|a)},
\end{equation}
and we can use this to evaluate
the reversibility of the state change as
\begin{equation}
  R(m) = \sum_a p(a|m)\,R(m,a).
\label{eq:defReversibility}
\end{equation}
This also quantifies the degree of disturbance
caused when a measurement $\{\hat{M}_m\}$
yields the outcome $m$, and satisfies
\begin{equation}
 0\le R(m) \le 1.
\end{equation}
Again, note that $R(m)$ relates to a single outcome,
unlike
\begin{equation}
R=\sum_m p(m)\,R(m),
\label{eq:avgR}
\end{equation}
which was discussed in Refs.~\cite{KoaUed99,CheLee12}.

As an important example, we consider
a diagonal measurement operator $\hat{M}^{(d)}_{k,l}(\lambda)$
with diagonal elements
\begin{equation}
  \underbrace{1,1,\ldots,1}_{k},
  \underbrace{\lambda,\lambda,\ldots,\lambda}_{l},
  \underbrace{0,0,\ldots,0}_{d-k-l}
\end{equation}
for $k=1,2,\ldots,d-1$ and $l=1,2,\ldots,d-k$,
with a parameter $\lambda$ satisfying $0\le\lambda\le1$.
In an orthonormal basis $\{\ket{i}\}$ with $i=1,2,\ldots,d$,
the measurement operator $\hat{M}^{(d)}_{k,l}(\lambda)$ can be written as
\begin{equation}
\hat{M}^{(d)}_{k,l}(\lambda)=\sum_{i=1}^{k} \ket{i}\bra{i}
                       +\sum_{i=k+1}^{k+l} \lambda \ket{i}\bra{i}.
\label{eq:measurementOp}
\end{equation}
The information that was yielded and disturbance that was caused
by this operator
can be quantified in terms of $I(m)$, $G(m)$, $F(m)$, and $R(m)$,
given by Eqs.~(\ref{eq:defInformation}), (\ref{eq:defEstimation}),
(\ref{eq:defFidelity}), and (\ref{eq:defReversibility})
as functions of the parameter $\lambda$.
Using the general formula derived in Ref.~\cite{Terash15},
$I(m)$ can be calculated to be
\begin{align}
 &I(m) = \log_2d
  -\frac{1}{\ln2}\Bigl[\eta(d)- 1\Bigr]
\notag\\
&\qquad\qquad
-\log_2\left(k+l\lambda^2\right)
+\frac{1}{k+l\lambda^2} J,
\label{eq:informationEx}
\end{align}
where $J$ is given by
\begin{align}
 & J =
(-1)^l \sum_{n=0}^{k-1}\binom{k+l-n-2}{l-1}\,
     \frac{a^{(k+l)}_n}{(\lambda^2-1)^{k+l-n-1}}
\notag\\
&\quad
+(-1)^k\sum_{n=0}^{l-1}\binom{k+l-n-2}{k-1}\,
     \frac{c^{(k+l)}_n(\lambda)}{(1-\lambda^2)^{k+l-n-1}},
\label{eq:dangerousEx}
\end{align}
with coefficients
\begin{align}
a^{(j)}_n
 &= \frac{1}{\ln2}\binom{j}{n}
       \Bigl[\eta(j)- \eta(j-n)\Bigr], \label{eq:defAjn} \\
c^{(j)}_n(\lambda)
 &= \lambda^{2(j-n)}
   \left[\binom{j}{n}\log_2 \lambda^2+a^{(j)}_n\right]
\end{align}
for $n=0,1,\ldots,j$.
Likewise, $G(m)$, $F(m)$, and $R(m)$ can be
calculated to be~\cite{Terash15}
\begin{align}
 G(m) &= \frac{1}{d+1}\left(1+\frac{1}{k+l\lambda^2}\right),
         \label{eq:estimationEx} \\
 F(m) &= \frac{1}{d+1}\left[1+\frac{(k+l\lambda)^2}
                          {k+l\lambda^2}\right],
         \label{eq:fidelityEx} \\
 R(m) &= d\left(\frac{\lambda^2}{k+l\lambda^2}\right)\,\delta_{d,(k+l)}.
         \label{eq:reversibilityEx}
\end{align}

\begin{figure}
\begin{center}
\includegraphics[scale=0.52]{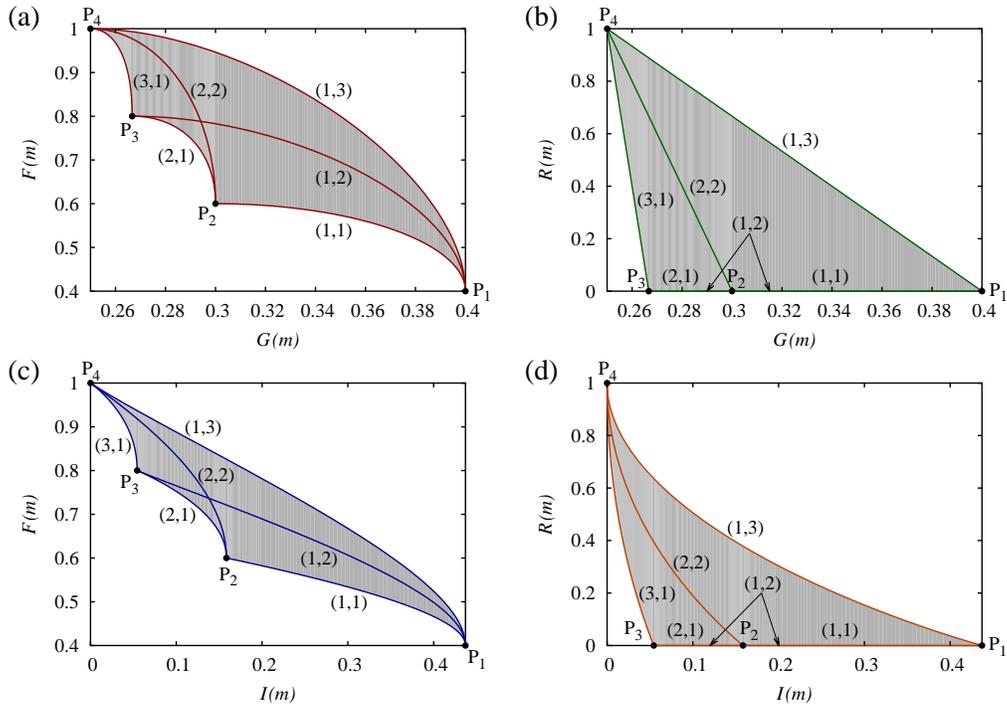}
\end{center}
\caption{\label{fig1}
Four allowed regions for information versus disturbance for $d=4$:
(a) estimation fidelity $G(m)$ versus operation fidelity $F(m)$;
(b) estimation fidelity $G(m)$ versus physical reversibility $R(m)$;
(c) information gain $I(m)$ versus operation fidelity $F(m)$; and
(d) information gain $I(m)$ versus physical reversibility $R(m)$.
}
\end{figure}%
The measurement operator $\hat{M}^{(d)}_{k,l}(\lambda)$
is very important
for obtaining the allowed regions in the information--disturbance planes
by plotting all physically possible measurement operators.
We consider four different allowed regions, based on using
$I(m)$ or $G(m)$ to quantify the information and
$F(m)$ or $R(m)$ to quantify the disturbance.
Figure~\ref{fig1} shows these four allowed regions
for $d=4$ in gray~\cite{Terash16},
where the lines $(k,l)$ correspond to $\hat{M}^{(d)}_{k,l}(\lambda)$
with $0\le\lambda\le1$
and the P${}_r$'s denote
the points corresponding to the projective measurement operator of rank $r$:
\begin{equation}
\hat{P}^{(d)}_{r}=\sum_{i=1}^{r} \ket{i}\bra{i}.
\end{equation}
Clearly, $\hat{M}^{(d)}_{k,l}(0)=\hat{P}^{(d)}_{k}$,
$\hat{M}^{(d)}_{k,l}(1)=\hat{P}^{(d)}_{k+l}$,
and $\hat{P}^{(d)}_{d}=\hat{I}$.
Thus, the line $(k,l)$ connects P${}_k$ to P${}_{k+l}$
and the point P${}_d$ is at the top left corner of the plot.
In Fig.~\ref{fig1},
the upper boundaries of the allowed regions
consist of the lines $(1,d-1)$
corresponding to $\hat{M}^{(d)}_{1,d-1}(\lambda)$,
whereas the lower boundaries consist of the lines $(k,1)$
corresponding to $\hat{M}^{(d)}_{k,1}(\lambda)$ for $k=1,2,\ldots,d-1$.
Therefore,
to find the values of the slopes and curvatures of the boundaries,
we need to calculate the first and second derivatives of
the disturbance with respect to information
for $\hat{M}^{(d)}_{k,l}(\lambda)$.

The above allowed regions were obtained by considering
ideal measurements, as in Eq.~(\ref{eq:postState}),
with optimal estimates for $G(m)$.
Unfortunately, the lower boundaries can be violated by
non-ideal measurements,
which yield mixed post-measurement states
due to classical noise,
or non-optimal estimates,
which make suboptimal choices for $\ket{\varphi(m)}$.
Here, we ignore such non-quantum effects
in order to focus on the quantum nature of measurement.

\section{\label{sec:derivativePara}Derivatives with respect to $\lambda^2$}
To calculate the derivative of
the disturbance with respect to information
for $\hat{M}^{(d)}_{k,l}(\lambda)$,
we first consider the derivatives of the information and disturbance
with respect to the parameter $\lambda^2$.
For simplicity, we focus on derivatives with respect to $\lambda^2$
rather than $\lambda$ itself.
These derivatives are straightforward to calculate
because the information and the disturbance are
expressed as functions of $\lambda=\sqrt{\lambda^2}$ in
Eqs.~(\ref{eq:informationEx}), (\ref{eq:estimationEx}),
(\ref{eq:fidelityEx}), and (\ref{eq:reversibilityEx}).

\begin{figure}
\begin{center}
\includegraphics[scale=0.6]{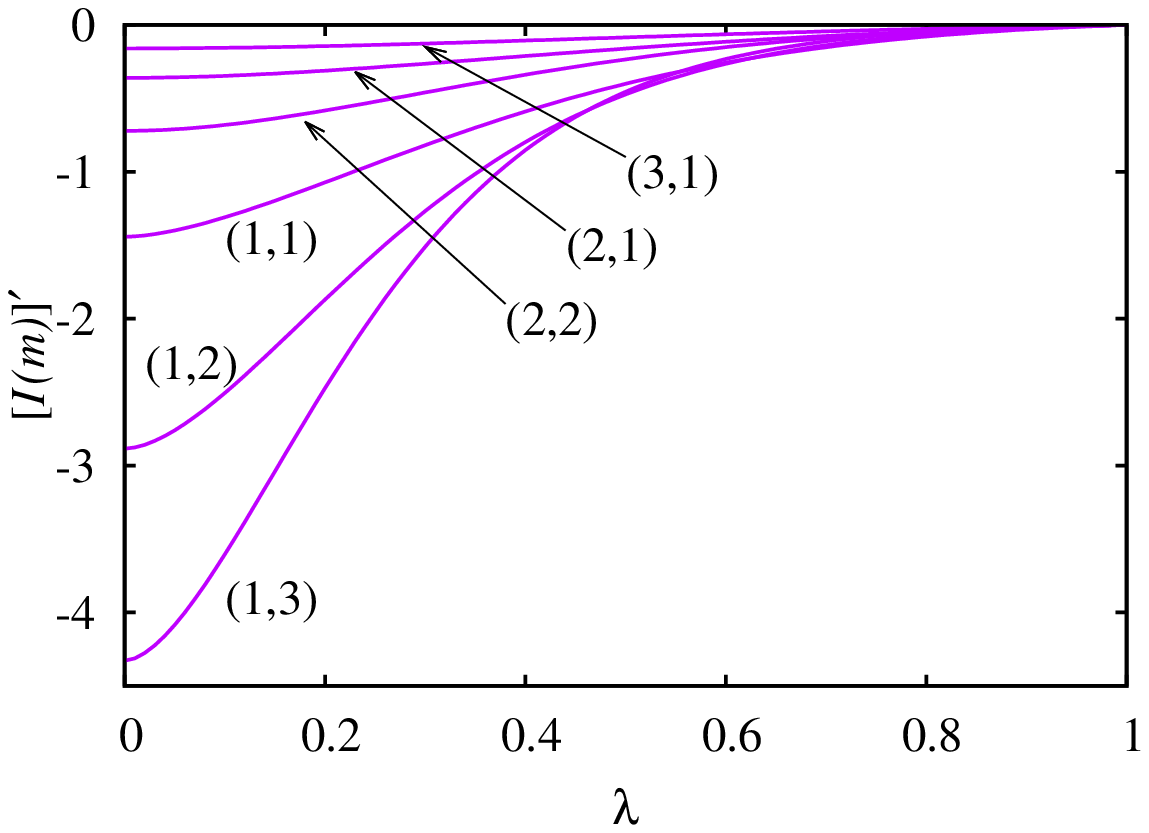}
\end{center}
\caption{\label{fig2}
First derivative of $I(m)$ with respect to $\lambda^2$
as a function of $\lambda$,
for $\hat{M}^{(d)}_{k,l}(\lambda)$ with $d=4$, for various $(k,l)$.}
\end{figure}

\begin{figure}
\begin{center}
\includegraphics[scale=0.6]{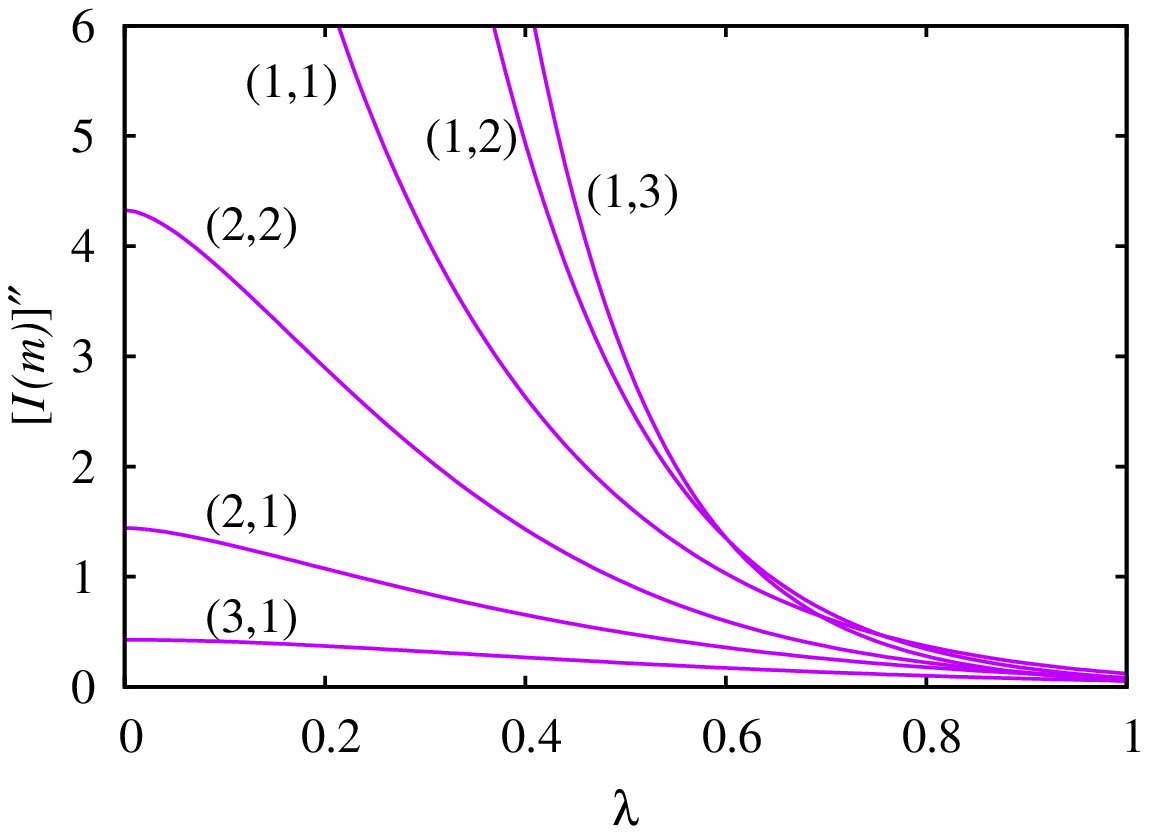}
\end{center}
\caption{\label{fig3}
Second derivative of $I(m)$ with respect to $\lambda^2$
as a function of $\lambda$,
for $\hat{M}^{(d)}_{k,l}(\lambda)$ with $d=4$, for various $(k,l)$.}
\end{figure}

However, the expression for 
the derivative of $I(m)$ is quite long.
This is due to the expression for $J$ given in Eq.~(\ref{eq:dangerousEx}).
From Eq.~(\ref{eq:informationEx}),
the first derivative of $I(m)$ is
\begin{align}
 [I(m)]' &= -\frac{1}{\ln2}\left(\frac{l}{k+l\lambda^2}\right) \notag \\
         &\qquad -\frac{l}{(k+l\lambda^2)^2} J +\frac{1}{k+l\lambda^2} J',
\label{eq:dIdl}
\end{align}
where primes represent derivatives with respect to $\lambda^2$.
The first derivative of $J$ can be written as
\begin{align}
 &J'=
(-1)^l\sum_{n=0}^{k-1}\binom{k+l-n-1}{l}\,
     \frac{-l a^{(k+l)}_n}{(\lambda^2-1)^{k+l-n}}
\notag\\
&\;\;\;
+(-1)^k\sum_{n=0}^{l-1}\binom{k+l-n-1}{k}\,
     \frac{k c^{(k+l)}_n(\lambda)}{(1-\lambda^2)^{k+l-n}}
\notag\\
&\;\;\;
+(-1)^k\sum_{n=0}^{l-1}\binom{k+l-n-2}{k-1}\,
     \frac{(n+1)\,c^{(k+l)}_{n+1}(\lambda)}{(1-\lambda^2)^{k+l-n-1}}
\label{eq:dJdl}
\end{align}
because
\begin{equation}
 [c^{(j)}_n(\lambda)]' = (n+1)\,c^{(j)}_{n+1}(\lambda).
\end{equation}
Figure~\ref{fig2} shows $[I(m)]'$
as a function of $\lambda$ for $d=4$, for various $(k,l)$.
From this, we can observe that $[I(m)]'\le0$.
In addition, the second derivative of $I(m)$ is
\begin{align}
 [I(m)]''
  &=\frac{1}{\ln2}\left[\frac{l^2}{(k+l\lambda^2)^2}\right]
    +\frac{2l^2}{(k+l\lambda^2)^3} J \notag \\
  &\qquad -\frac{2l}{(k+l\lambda^2)^2} J' +\frac{1}{k+l\lambda^2}J'',
\label{eq:d2Idl2}
\end{align}
and the second derivative of $J$ can be written as
\begin{align}
 &J'' =
 (-1)^{l} \sum_{n=0}^{k-1}\binom{k+l-n}{l+1}\,
    \frac{l(l+1) a^{(k+l)}_n}{(\lambda^2-1)^{k+l-n+1}}
\notag\\
&\;\;\;
+(-1)^k\sum_{n=0}^{l-1}\binom{k+l-n}{k+1}\,
   \frac{k(k+1) c^{(k+l)}_n(\lambda)}{(1-\lambda^2)^{k+l-n+1}}
\notag\\
&\;\;\;
+(-1)^k\sum_{n=0}^{l-1}\binom{k+l-n-1}{k}\,
    \frac{2k(n+1)\,c^{(k+l)}_{n+1}(\lambda)}{(1-\lambda^2)^{k+l-n}}
\notag\\
&\;\;\;
+(-1)^k\sum_{n=0}^{l-1}\binom{k+l-n-2}{k-1}\,
     \frac{(n+2)(n+1)\,c^{(k+l)}_{n+2}(\lambda)}{(1-\lambda^2)^{k+l-n-1}}.
\label{eq:d2Jdl2}
\end{align}
Figure~\ref{fig3} shows $[I(m)]''$
as a function of $\lambda$ for $d=4$, for various $(k,l)$.
From this, we can observe that $[I(m)]''>0$.

As shown in Appendix \ref{sec:limit0},
at $\lambda=0$,
$J$ and its derivatives become
\begin{gather}
 \lim_{\lambda\to 0} J=a^{(k)}_{k-1}, \quad
 \lim_{\lambda\to 0} J'=l a^{(k-1)}_{k-1}, \notag \\
 \lim_{\lambda\to 0} J''=
  \begin{cases}
      l(l+1) a^{(k-2)}_{k-1} & \mbox{(if $k\ge2$)} \\[5pt]
      +\infty & \mbox{(if $k=1$)},
   \end{cases}
  \label{eq:dangerAt0}
\end{gather}
where $a^{(j)}_{j+1}$ is given by
\begin{equation}
 a^{(j)}_{j+1} =\frac{1}{(j+1)\ln2}
\label{eq:defAjn2}
\end{equation}
instead of Eq.~(\ref{eq:defAjn}).
Here, $J''$ in Eq.~(\ref{eq:dangerAt0}) diverges for $k=1$
because
\begin{equation}
   \lim_{\lambda\to 0} c^{(j)}_j(\lambda)
    = \lim_{\lambda\to 0} \log_2\lambda^2+a^{(j)}_{j},
\label{eq:cjjAt0}
\end{equation}
which appears in the last sum of Eq.~(\ref{eq:d2Jdl2}) when $n=l-1$
if $k=1$.
The derivatives of $I(m)$ at $\lambda=0$ are thus
\begin{gather}
 \lim_{\lambda\to 0}\,[I(m)]'  = -\frac{l}{k^2\ln2}, \\
 \lim_{\lambda\to 0}\,[I(m)]'' =
   \begin{cases}
     \frac{l(k^2+3kl-2l)}{k^3(k-1)\ln2}
           & \mbox{(if $k\ge2$)} \\[5pt]
      +\infty & \mbox{(if $k=1$)}.
   \end{cases}
\label{eq:d2Idl2At0}
\end{gather}
Similarly, at $\lambda=1$,
$J$ and its derivatives become
\begin{gather}
 \lim_{\lambda\to 1} J=a^{(k+l)}_{k+l-1},   \quad
 \lim_{\lambda\to 1} J'=l a^{(k+l)}_{k+l}, \notag \\
 \lim_{\lambda\to 1} J''=l(l+1) a^{(k+l)}_{k+l+1},
\label{eq:dangerAt1}
\end{gather}
as shown in Appendix \ref{sec:limit1},
in which case the derivatives of $I(m)$ are
\begin{gather}
 \lim_{\lambda\to 1}\,[I(m)]'  = 0, \label{eq:dIdlAt1} \\
 \lim_{\lambda\to 1}\,[I(m)]'' = \frac{kl}{(k+l)^2(k+l+1)\ln2}.
  \label{eq:d2Idl2At1}
\end{gather}

Likewise, from Eqs.~(\ref{eq:estimationEx}), (\ref{eq:fidelityEx}),
and (\ref{eq:reversibilityEx}),
the first derivatives of $G(m)$, $F(m)$, and $R(m)$ are
\begin{align}
 [G(m)]' &= -\frac{l}{d+1}
     \left[\frac{1}{(k+l\lambda^2)^2}\right],
             \label{eq:dGdl} \\
 [F(m)]' &= \frac{kl}{d+1}
     \left[\frac{(1-\lambda)(k+l\lambda)}
                {\lambda(k+l\lambda^2)^2}\right],
             \label{eq:dFdl} \\
 [R(m)]' &= kd
    \left[\frac{1}{(k+l\lambda^2)^2} \right]\delta_{d,(k+l)},
             \label{eq:dRdl}
\end{align}
respectively.
These satisfy $[G(m)]'<0$, $[F(m)]'\ge0$, and $[R(m)]'\ge0$.
Note that $[R(m)]'$ is proportional to $[G(m)]'$
with a non-positive proportionality constant, i.e.,
$[R(m)]'=\alpha[G(m)]'$ with
\begin{equation}
\alpha=-\frac{kd(d+1)}{l}\delta_{d,(k+l)}.
\label{eq:propConst}
\end{equation}
In addition,
the second derivatives of $G(m)$, $F(m)$, and $R(m)$ are
\begin{align}
 [G(m)]'' &= \frac{2l^2}{d+1}
    \left[\frac{1}{(k+l\lambda^2)^3}\right],
             \label{eq:d2Gdl2} \\
 [F(m)]'' &= -\frac{kl}{2(d+1)}  \notag \\
    &\quad {}\times\left[
    \frac{(k+l\lambda^2)^2+4l\lambda^2(1-\lambda)(k+l\lambda)}
     {\lambda^3(k+l\lambda^2)^3}\right],
             \label{eq:d2Fdl2} \\
 [R(m)]'' &=  -2kld
     \left[\frac{1}{(k+l\lambda^2)^3}\right]\delta_{d,(k+l)},
             \label{eq:d2Rdl2}
\end{align}
respectively.
These satisfy $[G(m)]''>0$, $[F(m)]''<0$, and $[R(m)]''\le0$,
and $[R(m)]''$ is proportional to $[G(m)]''$
with the same proportionality constant $\alpha$,
given in Eq.~(\ref{eq:propConst}).

\begin{table}
\caption{\label{tbl1}
Signs of the first and second derivatives of the information and disturbance
with respect to $\lambda^2$.}
\begin{center}
\begin{tabular}{cccc}
\hline\noalign{\smallskip}
               &          & \multicolumn{2}{c}{derivative} \\
\noalign{\smallskip}\cline{3-4}\noalign{\smallskip}
               & function & first & second  \\
\noalign{\smallskip}\hline\noalign{\smallskip}
  information  & $I(m)$   & $-$ & $+$ \\
               & $G(m)$   & $-$ & $+$ \\
\noalign{\smallskip}\hline\noalign{\smallskip}
  disturbance  & $F(m)$   & $+$ & $-$ \\
               & $R(m)$   & $+$ & $-$ \\
\noalign{\smallskip}\hline
\end{tabular}
\end{center}
\end{table}
The signs of the derivatives of $I(m)$, $G(m)$, $F(m)$, and $R(m)$
are summarized in Table~\ref{tbl1}.
These signs mean that
when $\lambda^2$ is increased,
$I(m)$ and $G(m)$ decrease while $F(m)$ and $R(m)$ increase.
This is a trade-off between the information and the disturbance
for $\hat{M}^{(d)}_{k,l}(\lambda)$.

\section{\label{sec:derivativeInfo}Derivatives with respect to information}
Using the derivatives of the information and disturbance
with respect to $\lambda^2$,
we can now calculate the derivative of the disturbance
with respect to information for $\hat{M}^{(d)}_{k,l}(\lambda)$.
Let $f$ and $g$ be arbitrary functions of $\lambda$.
Given the derivatives of $f$ and $g$ with respect to $\lambda^2$,
the first and second derivatives of $f$ with respect to $g$ are
\begin{equation}
 \frac{df}{dg} = \frac{f'}{g'}, \quad
 \frac{d^2f}{dg^2} = \frac{f'' g'-f' g''}{(g')^3}.
\end{equation}
The same results can be obtained
using derivatives with respect to $\lambda$.

\begin{figure}
\begin{center}
\includegraphics[scale=0.52]{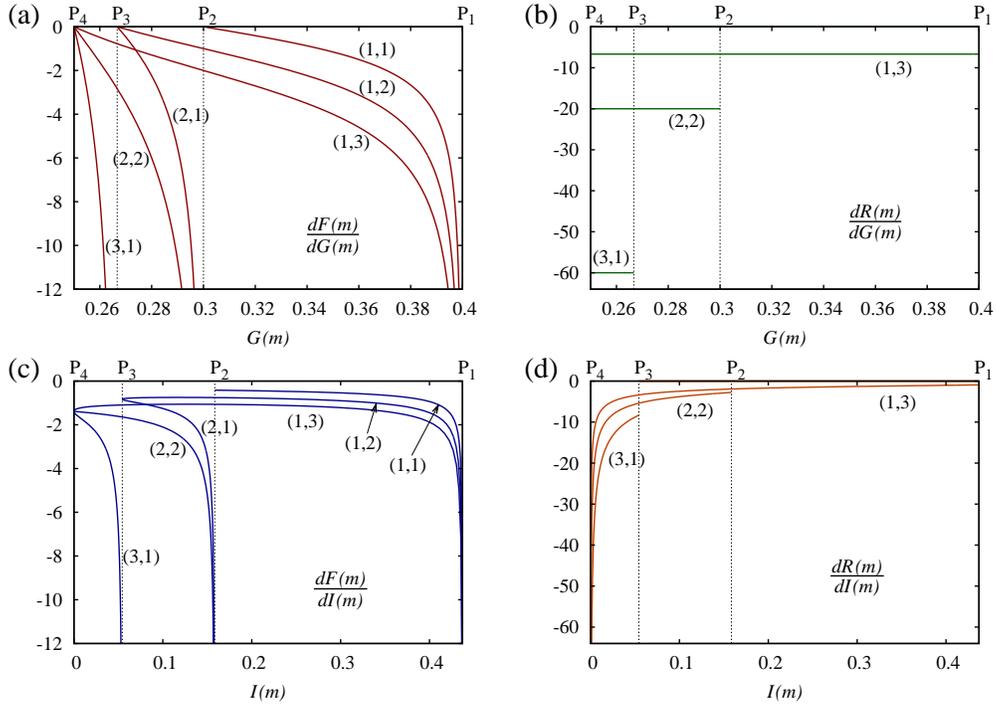}
\end{center}
\caption{\label{fig4}
First derivatives of the disturbance with respect to information
for $d=4$, for the four information--disturbance pairs:
(a) estimation fidelity $G(m)$ and operation fidelity $F(m)$;
(b) estimation fidelity $G(m)$ and physical reversibility $R(m)$;
(c) information gain $I(m)$ and operation fidelity $F(m)$; and
(d) information gain $I(m)$ and physical reversibility $R(m)$.
}
\end{figure}%
\begin{figure}
\begin{center}
\includegraphics[scale=0.52]{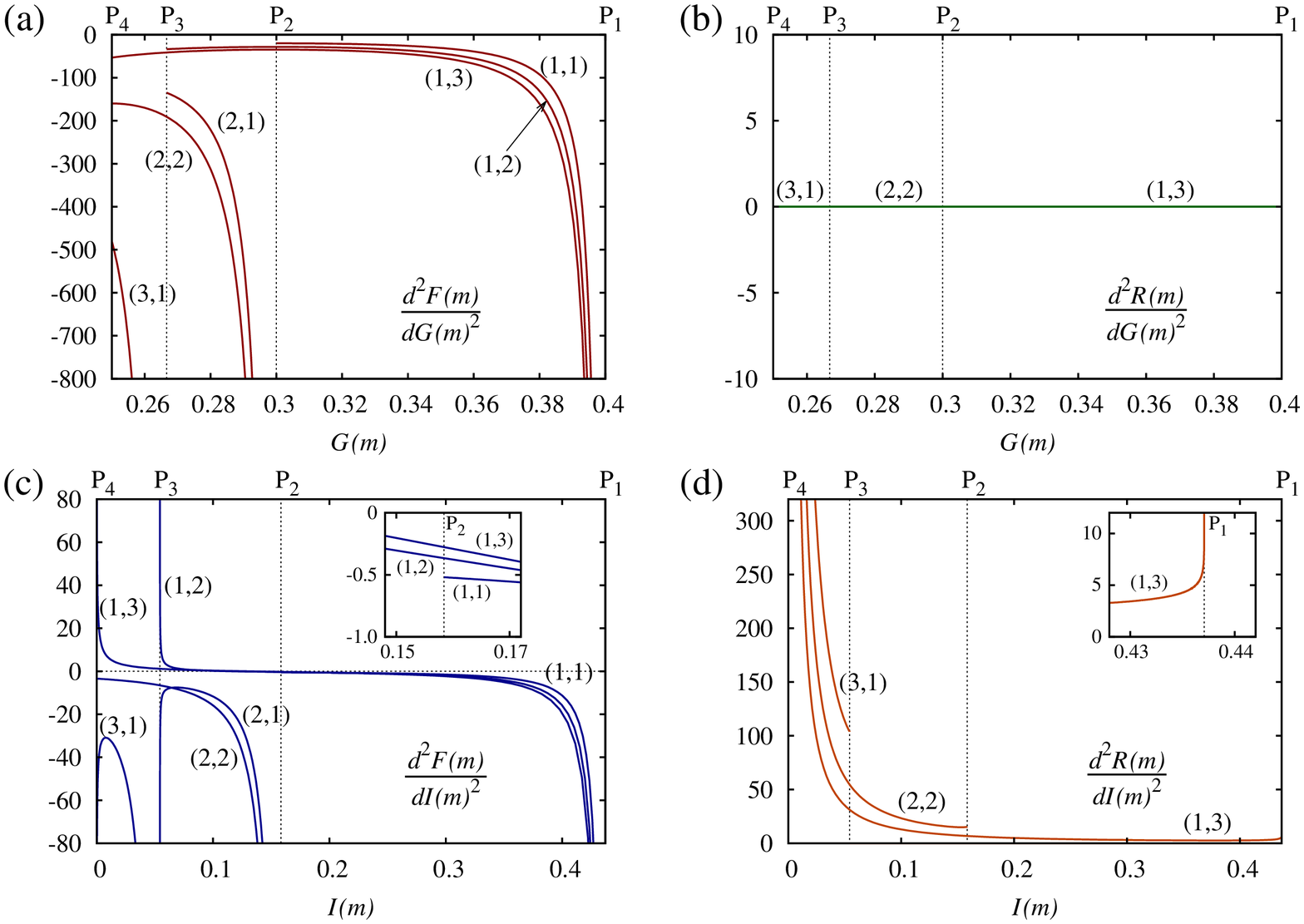}
\end{center}
\caption{\label{fig5}
Second derivatives of the disturbance with respect to information
for $d=4$, for the four information--disturbance pairs:
(a) estimation fidelity $G(m)$ and operation fidelity $F(m)$;
(b) estimation fidelity $G(m)$ and physical reversibility $R(m)$;
(c) information gain $I(m)$ and operation fidelity $F(m)$; and
(d) information gain $I(m)$ and physical reversibility $R(m)$.
}
\end{figure}%
From Eqs.~(\ref{eq:dGdl}), (\ref{eq:dFdl}), (\ref{eq:d2Gdl2}),
and (\ref{eq:d2Fdl2}),
the first and second derivatives of $F(m)$ with respect to $G(m)$
can be calculated to be
\begin{align}
 \frac{dF(m)}{dG(m)} &=
    -k\left[\frac{(1-\lambda)(k+l\lambda)}{\lambda}\right],
   \label{eq:dFdG} \\
 \frac{d^2F(m)}{dG(m)^2} &=
    -\frac{k(d+1)}{2l}\left[\frac{(k+l\lambda^2)^3}{\lambda^3}\right].
   \label{eq:d2FdG2}
\end{align}
Figures \ref{fig4}(a) and \ref{fig5}(a) show these derivatives
as functions of $G(m)$ [Eq.~(\ref{eq:estimationEx})]
for $d=4$, for various $(k,l)$.
Because $\lambda=0$ corresponds to P${}_{k}$ and
$\lambda=1$ corresponds to P${}_{k+l}$ for
the lines $(k,l)$ in Fig.~\ref{fig1},
the derivatives become
\begin{equation}
 \lim_{\lambda\to 0}\, \frac{dF(m)}{dG(m)} = -\infty, \quad
 \lim_{\lambda\to 0}\, \frac{d^2F(m)}{dG(m)^2} =  -\infty
\end{equation}
at P${}_{k}$ and
\begin{equation}
 \lim_{\lambda\to 1}\, \frac{dF(m)}{dG(m)} = 0, \quad
 \lim_{\lambda\to 1}\, \frac{d^2F(m)}{dG(m)^2} = -\frac{k(k+l)^3(d+1)}{2l}
\end{equation}
at P${}_{k+l}$.
The first derivative of $F(m)$ with respect to $G(m)$ [Eq.~(\ref{eq:dFdG})]
is non-positive and
the second derivative [Eq.~(\ref{eq:d2FdG2})] is negative,
which means that
all the lines $(k,l)$ in Fig.~\ref{fig1}(a) are
monotonically-decreasing convex curves.

In contrast,
from Eqs.~(\ref{eq:dGdl}), (\ref{eq:dRdl}), (\ref{eq:d2Gdl2}),
and (\ref{eq:d2Rdl2}),
the first and second derivatives of $R(m)$ with respect to $G(m)$
are constant:
\begin{align}
 \frac{dR(m)}{dG(m)} &=
    -\frac{kd(d+1)}{l},  \label{eq:dRdG} \\
 \frac{d^2R(m)}{dG(m)^2} &= 0  \label{eq:d2RdG2}
\end{align}
if $k+l=d$, and both derivatives are zero if $k+l\neq d$.
Figures \ref{fig4}(b) and \ref{fig5}(b) show
these derivatives as functions of $G(m)$ for $d=4$, for various $(k,l)$
satisfying $k+l=d$.
The first derivative of $R(m)$ with respect to $G(m)$ [Eq.~(\ref{eq:dRdG})]
is negative and
the second derivative [Eq.~(\ref{eq:d2RdG2})] is zero,
which means that all the lines $(k,l)$ in Fig.~\ref{fig1}(b) are
monotonically-decreasing straight lines.

Similarly,
from Eqs.~(\ref{eq:dIdl}), (\ref{eq:d2Idl2}), (\ref{eq:dFdl}),
and (\ref{eq:d2Fdl2}),
the first and second derivatives of $F(m)$ with respect to $I(m)$ are
\begin{align}
 \frac{dF(m)}{dI(m)} &= \frac{[F(m)]'}{[I(m)]'}, \label{eq:dFdI}  \\
 \frac{d^2F(m)}{dI(m)^2} &= \frac{[F(m)]''[I(m)]'-[F(m)]'[I(m)]''}
           {\left\{[I(m)]'\right\}^3}. \label{eq:d2FdI2}
\end{align}
Figures \ref{fig4}(c) and \ref{fig5}(c) show these derivatives 
as functions of $I(m)$ [Eq.~(\ref{eq:informationEx})]
for $d=4$, for various $(k,l)$.
At P${}_{k}$, they become
\begin{equation}
 \lim_{\lambda\to 0}\, \frac{dF(m)}{dI(m)} = -\infty, \quad
 \lim_{\lambda\to 0}\, \frac{d^2F(m)}{dI(m)^2} = -\infty
\end{equation}
because
\begin{equation}
\lim_{\lambda\to 0}\,[F(m)]' = \infty, \quad
 \lim_{\lambda\to 0}\, [F(m)]'' = -\infty.
\end{equation}
Note that the numerator of Eq.~(\ref{eq:d2FdI2})
goes to positive infinity as $\lambda\to0$
when $[I(m)]'<0$
because $[F(m)]''$ diverges faster than $[F(m)]'$.
In contrast, in the limit as $\lambda\to1$,
Eqs.~(\ref{eq:dFdI}) and (\ref{eq:d2FdI2})
yield the indeterminate form $0/0$
due to Eq.~(\ref{eq:dIdlAt1}) and
\begin{equation}
  \lim_{\lambda\to 1}\,[F(m)]' =0.
\label{eq:dFdlAt1}
\end{equation}
However,
by applying L'H{\^o}pital's rule and considering higher derivatives,
we can find that
\begin{gather}
 \lim_{\lambda\to 1}\, \frac{dF(m)}{dI(m)}
  = -\frac{(k+l)(k+l+1)\ln2}{2(d+1)}, \label{eq:dFdIAt1} \\
 \lim_{\lambda\to 1}\, \frac{d^2F(m)}{dI(m)^2}
  = \begin{cases}
      +\infty & \mbox{(if $k<l$)} \\[5pt]
      -\frac{k(2k+1)^3}{(2k+3)(d+1)} (\ln2)^2 & \mbox{(if $k=l$)} \\[5pt]
      -\infty  & \mbox{(if $k>l$)}
   \end{cases}
 \label{eq:d2FdI2At1}
\end{gather}
at P${}_{k+l}$, as shown in Appendix \ref{sec:LHopital}.
The first derivative of $F(m)$ with respect to $I(m)$
[Fig.~\ref{fig4}(c)] is negative,
and the second derivative [Fig.~\ref{fig5}(c)] is always negative if $k\ge l$
but can be positive near P${}_{k+l}$ if $k<l$.
This means that
the lines $(k,l)$ in Fig.~\ref{fig1}(c) are
monotonically-decreasing convex curves if $k\ge l$
but monotonically-decreasing S-shaped curves if $k<l$.
In particular,
even though it is difficult to see from Fig.~\ref{fig1}(c),
the upper boundary $(1,d-1)$ has a slight dent near P${}_d$
when $d\ge3$~\cite{Terash16}.

Finally, from Eqs.~(\ref{eq:dIdl}), (\ref{eq:d2Idl2}), (\ref{eq:dRdl}),
and (\ref{eq:d2Rdl2}),
the first and second derivatives of $R(m)$ with respect to $I(m)$ are
\begin{align}
 \frac{dR(m)}{dI(m)} &= \frac{[R(m)]'}{[I(m)]'}, \label{eq:dRdI} \\
 \frac{d^2R(m)}{dI(m)^2} &= \frac{[R(m)]''[I(m)]'-[R(m)]'[I(m)]''}
           {\left\{[I(m)]'\right\}^3}.      \label{eq:d2RdI2} 
\end{align}
Figures \ref{fig4}(d) and \ref{fig5}(d) show these derivatives
as functions of $I(m)$ for $d=4$, for various $(k,l)$
satisfying $k+l=d$.
(Both derivatives are zero if $k+l\neq d$.)
When $k+l=d$, they become
\begin{gather}
 \lim_{\lambda\to 0}\, \frac{dR(m)}{dI(m)} = -\frac{kd\ln2}{l}, \\
 \lim_{\lambda\to 0}\, \frac{d^2R(m)}{dI(m)^2}
  =\begin{cases}
     \frac{k^3}{k-1}\left(\frac{d\ln2}{l}\right)^2
           & \mbox{(if $k\ge2$)} \\[5pt]
      +\infty & \mbox{(if $k=1$)}
   \end{cases}
 \label{eq:dRdIAt0}
\end{gather}
at P${}_{k}$, and
\begin{equation}
 \lim_{\lambda\to 1}\, \frac{dR(m)}{dI(m)} = -\infty, \quad
 \lim_{\lambda\to 1}\, \frac{d^2R(m)}{dI(m)^2} = +\infty
 \label{eq:dRdIAt1}
\end{equation}
at P${}_{k+l}$.
In Eq.~(\ref{eq:dRdIAt0}), the second derivative diverges for $k=1$
because of the corresponding result in Eq.~(\ref{eq:d2Idl2At0}),
and the divergences seen in Eq.~(\ref{eq:dRdIAt1})
likewise come from Eq.~(\ref{eq:dIdlAt1}).
Note that
\begin{equation}
 \lim_{\lambda\to 1}\, \frac{1}{[I(m)]'} =-\infty
\label{eq:invdIdlAt1}
\end{equation}
because $[I(m)]'$ tends to zero from below,
as shown in Fig.~\ref{fig2}.
The first derivative of $R(m)$ with respect to $I(m)$
[Fig.~\ref{fig4}(d)] is negative
and the second derivative [Fig.~\ref{fig5}(d)] is positive,
which means that
all the lines $(k,l)$ in Fig.~\ref{fig1}(d) are
monotonically-decreasing concave curves.

\begin{table}
\caption{\label{tbl2}
Signs of the first and second derivatives of the disturbance
with respect to information.}
\begin{center}
\begin{tabular}{cccc}
\hline\noalign{\smallskip}
                &             & \multicolumn{2}{c}{derivative} \\
\noalign{\smallskip}  \cline{3-4}\noalign{\smallskip}
    disturbance & information & first & second  \\
\noalign{\smallskip}\hline\noalign{\smallskip}
       $F(m)$  &  $G(m)$  & $-$ & $-$ \\
       $R(m)$  &  $G(m)$  & $-$ & $0$ \\
       $F(m)$  &  $I(m)$  & $-$ & $\pm$ \\
       $R(m)$  &  $I(m)$  & $-$ & $+$ \\
\noalign{\smallskip}\hline
\end{tabular}
\end{center}
\end{table}
The signs of the derivatives for
the four information--disturbance pairs
are summarized in Table~\ref{tbl2}.
All the first derivatives have negative signs,
which implies that
there is a trade-off between the information and the disturbance
for each of the four pairs.
In contrast,
the second derivatives have different signs,
which implies that
the optimal measurements are different
for each of the four pairs~\cite{Terash16}.

\section{\label{sec:conclude}Conclusion}
In this paper, we have obtained the first and second derivatives
of the disturbance with respect to information
for a class of quantum measurements
described by the measurement operator $\hat{M}^{(d)}_{k,l}(\lambda)$
[Eq.~(\ref{eq:measurementOp})].
When the measurement performed on a $d$-level system
in a completely unknown state yields a single outcome $m$,
the information is quantified by
the Shannon entropy $I(m)$ [Eq.~(\ref{eq:informationEx})] and
the estimation fidelity $G(m)$ [Eq.~(\ref{eq:estimationEx})],
while the disturbance is quantified by
the operation fidelity $F(m)$ [Eq.~(\ref{eq:fidelityEx})] and
the physical reversibility $R(m)$ [Eq.~(\ref{eq:reversibilityEx})].
In these four information--disturbance planes,
$\hat{M}^{(d)}_{k,l}(\lambda)$ with $0\le\lambda\le1$
corresponds to a line $(k,l)$, as shown in Fig.~\ref{fig1}.
In particular,
the lines $(1,d-1)$ and $(k,1)$
form the boundaries of the allowed regions
obtained by plotting 
all physically possible measurement operators
in these planes~\cite{Terash16}.

The slope and curvature of each line $(k,l)$ are given by
the first and second derivatives of
the disturbance with respect to the information
for $\hat{M}^{(d)}_{k,l}(\lambda)$.
For these four information--disturbance pairs,
the first derivatives are given by
Eqs.~(\ref{eq:dFdG}), (\ref{eq:dRdG}), (\ref{eq:dFdI}), and (\ref{eq:dRdI})
(shown for $d=4$ in Fig.~\ref{fig4}),
while the second derivatives are given by
Eqs.~(\ref{eq:d2FdG2}), (\ref{eq:d2RdG2}), (\ref{eq:d2FdI2}),
and (\ref{eq:d2RdI2})
(shown for $d=4$ in Fig.~\ref{fig5}).
For the derivative of $F(m)$ with respect to $G(m)$,
all the lines $(k,l)$ in Fig.~\ref{fig1}(a)
are monotonically-decreasing convex curves,
because the first and second derivatives are
non-positive and negative, respectively,
as shown in Figs.~\ref{fig4}(a) and \ref{fig5}(a).
For the derivative of $R(m)$ with respect to $G(m)$,
all the lines $(k,l)$ in Fig.~\ref{fig1}(b)
are monotonically-decreasing straight lines,
because the first and second derivatives are
negative and zero, respectively,
as shown in Figs.~\ref{fig4}(b) and \ref{fig5}(b).
For the derivative of $F(m)$ with respect to $I(m)$,
the lines $(k,l)$ in Fig.~\ref{fig1}(c) are
monotonically-decreasing convex curves if $k\ge l$
and monotonically-decreasing S-shaped curves if $k<l$,
because the first derivative is negative and
the second derivative is always negative if $k\ge l$
but can be positive near P${}_{k+l}$ if $k<l$,
as shown in Figs.~\ref{fig4}(c) and \ref{fig5}(c).
Finally, for the derivative of $R(m)$ with respect to $I(m)$,
all the lines $(k,l)$ in Fig.~\ref{fig1}(d)
are monotonically-decreasing concave curves,
because the first and second derivatives are
negative and positive, respectively,
as shown in Figs.~\ref{fig4}(d) and \ref{fig5}(d).
See also Table~\ref{tbl2} for a summary of
the signs of the derivatives.

Based on these results, we can see that
the boundaries $(1,d-1)$ and $(k,1)$
of the allowed regions have non-positive slopes
for all four information--disturbance pairs,
indicating that
there is a trade-off between the information and the disturbance
for measurements on their boundaries.
When the information is increased by moving along a boundary,
the disturbance also increases, decreasing $F(m)$ and $R(m)$.
In addition, the rate of change of the disturbance with respect to
information is given by the boundary's slope.
For example,
if $G(m)$ is increased by $\Delta G(m)$,
$F(m)$ decreases by about
\begin{equation}
  \Delta F(m)= \left|\frac{dF(m)}{dG(m)}\right|\Delta G(m).
\end{equation}
Figure \ref{fig4}(a) shows that
$\left|dF(m)/dG(m)\right|$
is infinitely large near P${}_{1}$,
but almost zero near P${}_{d}$.

In contrast,
the curvatures of the boundaries $(1,d-1)$ and $(k,1)$
for the four information--disturbance pairs have
different signs.
This means that
the allowed regions are extended in different ways
when the information and disturbance
are averaged over all possible outcomes,
as with $I$, $G$, $F$, and $R$,
given by Eqs.~(\ref{eq:avgI}), (\ref{eq:avgG}),
(\ref{eq:avgF}), and (\ref{eq:avgR}),
because the allowed regions for the average values
are the convex hulls of those for a single outcome~\cite{Terash16}.
The upper boundaries
of the allowed regions for the average values
correspond to the optimal measurements that saturate
the upper information bounds for a given disturbance.
Consequently, the optimal measurements are different
for each of the four information--disturbance pairs~\cite{Terash16}.

\appendix
\section*{Appendix}

\section{\label{sec:limit0}Limits as $\lambda\to0$}
Here, we show that the first and second derivatives of $J$
with respect to $\lambda^2$ are as given in Eq.~(\ref{eq:dangerAt0})
in the limit as $\lambda\to0$.
First, note that
\begin{equation}
 \lim_{\lambda\to 0} J= a^{(k)}_{k-1},
\end{equation}
as shown in Appendix C of Ref.~\cite{Terash15}.
The limits of these derivatives can also be shown in a similar way.

For example, at $\lambda=0$,
the first derivative of $J$ [Eq.~(\ref{eq:dJdl})] becomes
\begin{equation}
 \lim_{\lambda\to 0} J'=
\sum_{n=0}^{k-1}\binom{k+l-n-1}{l}\,
     (-1)^{k-n-1}l a^{(k+l)}_n,
\label{eq:dJdlAt0}
\end{equation}
because $c^{(j)}_n(0)=0$ if $n<j$.
This equation can be simplified by using the identity
\begin{equation}
\sum_{n=0}^{k-1}\binom{k+l-n-1}{l}\,
     (-1)^{k-n-1} a^{(k+l)}_n = a^{(k-1)}_{k-1},
\label{eq:id4dJdlAt0}
\end{equation}
which can be derived from
\begin{align}
 & \frac{1}{(1+\epsilon)^{l+1}}\left[\left(1+\epsilon\right)^{k+l} \log_2 
  \left(1+\epsilon\right)  \right]
    \notag \\
 & \qquad\qquad\quad = \left(1+\epsilon\right)^{k-1} \log_2 
  \left(1+\epsilon\right)
\label{eq:eq4dJdlAt0}
\end{align}
by expanding every factor as a Taylor series.
In other words, the first factor in Eq.~(\ref{eq:eq4dJdlAt0})
can be expanded using the generalized binomial theorem
\begin{equation}
\frac{1}
  {\left(1+\epsilon\right)^{j}}
 = \sum_{n=0}^{\infty}
    \binom{j-1+n}{j-1}\left(-1\right)^{n}\epsilon^n,
\label{eq:TaylorBinom}
\end{equation}
while the other factors
can be expanded in terms of coefficients $\{a^{(j)}_n\}$~\cite{Terash15},
\begin{equation}
\left(1+\epsilon\right)^j \log_2 
  \left(1+\epsilon\right) =
\sum_{n=0}^{\infty} a^{(j)}_n\,\epsilon^n,
\label{eq:TaylorAjn}
\end{equation}
where the $a^{(j)}_{n}$'s
are given by Eq.~(\ref{eq:defAjn}) for $n=0,1,\ldots,j$
and by
\begin{equation}
 a^{(j)}_{n} =\frac{(-1)^{n-j-1}}{\ln2}\left[\frac{j!\,(n-j-1)!}{n!}\right]
\label{eq:defAjn3}
\end{equation}
for $n=j+1,j+2,\ldots$.
In particular, Eq.~(\ref{eq:defAjn3}) reduces to Eq.~(\ref{eq:defAjn2})
for $n=j+1$.
The identity in Eq.~(\ref{eq:id4dJdlAt0}) can then be proven
by substituting Eqs.~(\ref{eq:TaylorBinom}) and (\ref{eq:TaylorAjn})
into Eq.~(\ref{eq:eq4dJdlAt0})
and comparing the terms of order $\epsilon^{k-1}$ on both sides.
Substituting Eq.~(\ref{eq:id4dJdlAt0}) into Eq.~(\ref{eq:dJdlAt0}),
we find that $J'$ is
\begin{equation}
 \lim_{\lambda\to 0} J'= la^{(k-1)}_{k-1}
\end{equation}
at $\lambda=0$, as given in Eq.~(\ref{eq:dangerAt0}).

Similarly, the second derivative of $J$
[Eq.~(\ref{eq:d2Jdl2})] can be shown to be
\begin{equation}
 \lim_{\lambda\to 0} J''= l(l+1) a^{(k-2)}_{k-1}
\end{equation}
if $k\ge2$ by using the identity
\begin{equation}
 \sum_{n=0}^{k-1}\binom{k+l-n}{l+1}\,
    (-1)^{k-n-1}a^{(k+l)}_n = a^{(k-2)}_{k-1},
\end{equation}
which can be derived from the terms of order $\epsilon^{k-1}$ in
\begin{align}
 & \frac{1}{(1+\epsilon)^{l+2}}\left[\left(1+\epsilon\right)^{k+l} \log_2 
  \left(1+\epsilon\right)  \right]
   \notag \\
 &\qquad\qquad\quad
   = \left(1+\epsilon\right)^{k-2} \log_2 
  \left(1+\epsilon\right).
\end{align}
However, if $k=1$, $J''$ contains $c^{(l+1)}_{l+1}(\lambda)$,
which diverges in the limit as $\lambda\to0$,
as shown by Eq.~(\ref{eq:cjjAt0}).
By combining these results, we find that
$J''$ is given by Eq.~(\ref{eq:dangerAt0}) at $\lambda=0$.

\section{\label{sec:limit1}Limits as $\lambda\to1$}
Here, we show that the first and second derivatives of $J$
with respect to $\lambda^2$ are as given in Eq.~(\ref{eq:dangerAt1})
in the limit as $\lambda\to1$.
To find the derivatives at $\lambda=1$,
we first obtain the Taylor series for
$J$ around $\lambda=1$
by substituting $\lambda^2=1-\epsilon$ into Eq.~(\ref{eq:dangerousEx}):
\begin{equation}
  J=\sum_{n=0}^\infty j_n\, (-\epsilon)^n.
\end{equation}
Note that
the terms with negative powers of $\epsilon$
cancel each other out in this expansion
because $J$ is finite, even at $\lambda=1$~\cite{Terash15}.
The coefficients $\{j_n\}$ are related to 
the derivatives of $J$ at $\lambda=1$ by
\begin{equation}
  \lim_{\lambda\to 1} J =j_0,\quad
  \lim_{\lambda\to 1} J'=j_1,\quad
  \lim_{\lambda\to 1} J''=2j_2.
\label{eq:jn2dangerAt1}
\end{equation}
In Appendix C of Ref.~\cite{Terash15}, $j_0$ was shown to be
$a^{(k+l)}_{k+l-1}$, as given in Eq.~(\ref{eq:dangerAt1}),
and the other coefficients can be handled similarly.

For example, by applying Eq.~(\ref{eq:TaylorAjn}) to
$c^{(j)}_n\left(\sqrt{1-\epsilon}\right)$,
$j_1$ can be given as
\begin{align}
  j_1 &=\sum_{n=0}^{l-1}\binom{k+l-n-2}{k-1}\,(-1)^{l-n-1} \notag \\
    &\qquad\quad\times
     \left[\binom{k+l}{n}a^{(k+l-n)}_{k+l-n}+a^{(k+l)}_n \right].
\end{align}
The expression in the square brackets satisfies
\begin{equation}
  \binom{k+l}{n}a^{(k+l-n)}_{k+l-n}+a^{(k+l)}_n 
   =\binom{k+l}{n}\,a^{(k+l)}_{k+l},
\end{equation}
from Eq.~(\ref{eq:defAjn}).
By using the identity
\begin{equation}
\sum_{n=0}^{l-1}\,(-1)^{l-n-1} \,\binom{k+l-n-2}{k-1}\,
      \binom{k+l}{n}=l,
\end{equation}
which can be derived from the terms of order $\epsilon^{l-1}$ in
\begin{equation}
   \frac{1}{(1+\epsilon)^k}\left(1+\epsilon\right)^{k+l}
    =\left(1+\epsilon\right)^{l},
\end{equation}
we find that $j_1$ is
\begin{equation}
 j_1= la^{(k+l)}_{k+l}.
\end{equation}
Therefore, from Eq.~(\ref{eq:jn2dangerAt1}),
we see that $J'$ is given by Eq.~(\ref{eq:dangerAt1}) at $\lambda=1$.

Similarly, $j_2$ is given by
\begin{align}
  j_2 &=\sum_{n=0}^{l-1}\binom{k+l-n-2}{k-1}\,(-1)^{l-n-1} \notag \\
    &\qquad\quad\times
     \binom{k+l}{n}a^{(k+l-n)}_{k+l-n+1}.
\end{align}
From Eq.~(\ref{eq:defAjn2}), the last factor satisfies
\begin{equation}
  \binom{k+l}{n}a^{(k+l-n)}_{k+l-n+1}
   =\binom{k+l+1}{n}\,a^{(k+l)}_{k+l+1}.
\end{equation}
By using the identity
\begin{align}
 & \sum_{n=0}^{l-1}\,(-1)^{l-n-1} \,\binom{k+l-n-2}{k-1}\,
      \binom{k+l+1}{n}
  \notag \\
 & \qquad\qquad\qquad\qquad\qquad\qquad\quad
    =\frac{l(l+1)}{2},
\end{align}
which can be derived from the terms of order $\epsilon^{l-1}$ in
\begin{equation}
   \frac{1}{(1+\epsilon)^k}\left(1+\epsilon\right)^{k+l+1}
    =\left(1+\epsilon\right)^{l+1},
\end{equation}
we find that $j_2$ is
\begin{equation}
 j_2= \frac{l(l+1)}{2}a^{(k+l)}_{k+l+1}.
\end{equation}
Therefore, from Eq.~(\ref{eq:jn2dangerAt1}),
we see that $J''$ is given by Eq.~(\ref{eq:dangerAt1}) at $\lambda=1$.

In general, we can use a similar argument to find that $j_n$ is
\begin{equation}
  j_n = \binom{l-1+n}{l-1}a^{(k+l)}_{k+l-1+n},
\end{equation}
which shows that the $n$th derivative of $J$ at $\lambda=1$ is given by
\begin{equation}
 \lim_{\lambda\to 1} J^{(n)}=n!j_n
   =\frac{(l-1+n)!}{(l-1)!}a^{(k+l)}_{k+l-1+n}.
\label{eq:dnJdlnAt1}
\end{equation}

\section{\label{sec:LHopital}
Derivative calculations using L'H{\^o}pital's rule}
Here, we show that
the first and second derivatives of $F(m)$ with respect to $I(m)$
are as given in Eqs.~(\ref{eq:dFdIAt1}) and (\ref{eq:d2FdI2At1}),
respectively, in the limit as $\lambda\to1$.
We need to apply L'H{\^o}pital's rule to find these derivatives,
because Eqs.~(\ref{eq:dFdI}) and (\ref{eq:d2FdI2}) yield
the indeterminate form $0/0$ in the limit as $\lambda\to1$,
due to Eqs.~(\ref{eq:dIdlAt1}) and (\ref{eq:dFdlAt1}).
By applying L'H{\^o}pital's rule to Eq.~(\ref{eq:dFdI}),
we can give the first derivative as
\begin{equation}
 \lim_{\lambda\to 1}\frac{dF(m)}{dI(m)} =
   \lim_{\lambda\to 1} \frac{[F(m)]''}{[I(m)]''},
\end{equation}
which allows us to show Eq.~(\ref{eq:dFdIAt1}) based on
Eqs.~(\ref{eq:d2Idl2At1}) and (\ref{eq:d2Fdl2}).

Similarly,
by applying L'H{\^o}pital's rule to Eq.~(\ref{eq:d2FdI2}) twice,
we can give the second derivative as
\begin{align}
&\lim_{\lambda\to 1}\frac{d^2F(m)}{dI(m)^2}  \notag \\
&\quad=
  \lim_{\lambda\to 1} \frac{\left\{[F(m)]''[I(m)]'-[F(m)]'[I(m)]''\right\}''}
           {\left\{\left\{[I(m)]'\right\}^3\right\}''}.
\label{eq:d2FdI2LHopital}
\end{align}
This equation requires
the third derivatives of $I(m)$ and $F(m)$ at $\lambda=1$.
By differentiating Eqs.~(\ref{eq:d2Idl2}) and (\ref{eq:d2Fdl2})
and using Eq.~(\ref{eq:dnJdlnAt1}),
these can be calculated to be
\begin{align}
 \lim_{\lambda\to 1} [I(m)]''' &=
    -\frac{2kl(3kl+3l^2+k+5l)}{(k+l)^3(k+l+1)(k+l+2)\ln2}, \\
 \lim_{\lambda\to 1} [F(m)]''' &=
    \frac{3kl(k+3l)}{4(d+1)(k+l)^2}.
\end{align}
Then, we note that
the numerator of Eq.~(\ref{eq:d2FdI2LHopital}) can be written
as $(k-l)A$ with a positive constant $A$ at $\lambda=1$,
whereas its denominator goes to zero from below as $\lambda\to1$.
Therefore, from Eq.~(\ref{eq:invdIdlAt1}), we find that
Eq.~(\ref{eq:d2FdI2LHopital}) goes to positive infinity if $k<l$
and negative infinity if $k>l$, as given in Eq.~(\ref{eq:d2FdI2At1}),
but it still yields the indeterminate form $0/0$ if $k=l$.
However, by applying L'H{\^o}pital's rule once again in this case,
the second derivative can be given as
\begin{align}
&\lim_{\lambda\to 1}\frac{d^2F(m)}{dI(m)^2}  \notag \\
&\quad=
  \lim_{\lambda\to 1} \frac{\left\{[F(m)]''[I(m)]'-[F(m)]'[I(m)]''\right\}'''}
           {\left\{\left\{[I(m)]'\right\}^3\right\}'''}.
\label{eq:d2FdI2LHopital2}
\end{align}
If $k=l$,
from Eqs.~(\ref{eq:d2Idl2}), (\ref{eq:d2Fdl2}), and (\ref{eq:dnJdlnAt1}),
the fourth derivatives of $I(m)$ and $F(m)$ at $\lambda=1$
can be calculated to be
\begin{align}
 \lim_{\lambda\to 1} [I(m)]'''' &=
    \frac{3(12k+19)}
    {8(2k+1)(2k+3)\ln2}, \\
 \lim_{\lambda\to 1} [F(m)]'''' &=
    -\frac{39k}{16(d+1)}.
\end{align}
Substituting these derivatives into Eq.~(\ref{eq:d2FdI2LHopital2})
allows us to show Eq.~(\ref{eq:d2FdI2At1}) for $k=l$.

%\bibliographystyle{prsty}
%\bibliography{art,bk,paper}
%\end{document}

\end{document}